\documentclass[aps,prl,twocolumn,floatfix]{revtex4}
\usepackage{amsmath}
\usepackage{graphicx}
\usepackage{latexsym}
\usepackage{amsfonts}
\usepackage{amssymb}

\begin{document}

\newcommand{\bra}[1]{\ensuremath{\left< #1 \right|}}
\newcommand{\ket}[1]{\ensuremath{\left| #1 \right>}}

\title{Zero-area single photon pulses}

\author{L. S. Costanzo$^{1,2}$, A. S. Coelho$^{3}$, D. Pellegrino$^{2}$, M. S. Mendes$^{4}$, L. Acioli$^{4}$, K. N. Cassemiro$^{4}$, D. Felinto$^{4}$, A. Zavatta$^{1,2}$, and M. Bellini$^{1,2}$}

\affiliation{$^1$Istituto Nazionale di Ottica (INO-CNR), L.go E. Fermi 6, 50125 Florence, Italy\\
$^2$European Laboratory for non linear Spectroscopy (LENS) and Department of Physics, University of Firenze, 50019 Sesto Fiorentino, Florence, Italy\\
$^3$Departamento de Engenharia, Instituto Camillo Filho, 64049-230 Teresina, PI, Brazil\\
$^4$Departamento de Fisica, Universidade Federal de Pernambuco, 50670-901 Recife. PE, Brazil}

\begin{abstract}
Broadband single photons are usually considered not to couple efficiently to atomic gases because of the large mismatch in bandwidth. Contrary to this intuitive picture, here we demonstrate that the interaction of ultrashort single photons with a dense resonant atomic sample deeply modifies the temporal shape of their wavepacket mode without degrading their non-classical character, and effectively generates zero-area single-photon pulses. This is a clear signature of strong transient coupling between single broadband (THz-level) light quanta and atoms, with intriguing fundamental implications and possible new applications to the storage of quantum information. 
\end{abstract}
\date{\today}

\maketitle

Single photons are privileged carriers of quantum information because of their little interaction with the environment and among themselves. However, when it comes to storing and manipulating information, it would be useful for them to interact strongly with some atomic system in order to convert their quantum state into a stationary quantum state of matter~\cite{Lvovsky2009}.
Since atomic systems, either made of cold and ultracold atoms or of hot vapors, have absorption linewidths in the Hz to GHz range, the main road to enhance the atom-photon interaction has always been that of using sufficiently narrowband quantum photonic states, either produced in cavity-enhanced parametric down-conversion sources~\cite{Bao2008,Scholz2009,Wolfgramm2011,Zhang2011}, or directly from cold~\cite{Chaneliere2005,Thomson2006,Choi2008,Lettner2011} or hot~\cite{Eisaman2005,Mcrae2012} atomic samples. In general, ultrashort single photons with bandwidths much broader than the atomic bandwidth are therefore not considered useful for this task because they are thought to interact only very weakly with the atoms. This is not necessarily true.

Resonant interaction between ultrashort classical pulses and atomic media has long been investigated, together with some of its most peculiar effects. Two-photon transitions, for example, are well known to involve the whole bandwidth of an ultrashort pulse and to benefit considerably from the broad shaping of its spectral profile~\cite{Meshulach1998,Dudovich2001}. The formation of zero-area pulses is another spectacular consequence of the propagation of weak ultrashort pulses in the dispersive medium around an atomic resonance~\cite{Crisp1970,Rothenberg1984}.
Considering a laser pulse whose description in frequency space is initially given by $E(\omega,0)$, propagation through a distance $l$ in the resonant medium modifies it to:
\begin{equation}
E(\omega,l)=E(\omega,0) \exp \left[ \frac{-\alpha_0 l}{1-i(\omega-\omega_a)T_2}\right] ,
\label{eq_spmod}
\end{equation}
where $\alpha_0$ is the optical density of the medium, $\omega_a$ is the atomic resonance frequency, and $T_2$ is the upper level lifetime. This approximate expression, which considers two-level atoms and an effective single lorentzian profile for the resonance line of the sample is enough to convey all the main features of the phenomenon.

If the atomic transition is sufficiently narrow, the absorbed pulse energy can be almost negligible even in the case of high optical depths, but dispersion may still cause a dramatic re-shaping of the temporal pulse envelope. In accordance to the pulse area theorem~\cite{Arecchi1965,McCall1967,Allen1987}, the electric field amplitude of the pulse rapidly develops a series of lobes of alternating signs that make the pulse area approach zero. Once formed, a 0-area pulse is remarkably robust and propagates without further losses.

Although most studies concentrated on the effects of propagation on the shape of the optical pulse itself, it has also been pointed out that 0-area pulses can significantly enhance the transient excitation of the atoms~\cite{Dudovich2002}, even though the final excitation left in the medium after the passage of the pulse is negligible. This transient excitation, however, can be mapped into a final, non-negligible excitation of a different atomic level by means of a second ultrashort field acting on the excited state during the transient~\cite{Dudovich2002}, indicating its possible use for storage of information. All previous experimental studies of broadband pulse interaction with narrowband atomic ensembles were carried out with weak classical light pulses (see, for example,~\cite{Rothenberg1984,Matusovsky1996,Kallmann1999,Felinto2000}). However, the formation of 0-area single-photon (SP) pulses has also been predicted~\cite{Derouault2012} and viewed as a possible way to engineer quantum states using time-dependent effects and for time-domain quantum information processing~\cite{Petrosyan2013}. 

Here we present the first experimental demonstration of 0-area pulse formation within the wavepacket temporal mode of a single photon. We propagate ultrashort heralded photons with a broad bandwidth centered at 780 nm through a cell containing resonant hot rubidium vapor and find that, despite a negligibly small absorption, the transmitted SP wavepackets acquire the strong temporal modulation characteristic of 0-area pulses. We use our recently developed techniques of ultrafast, mode-selective, time-domain, homodyne detection~\cite{Polycarpou2012} to measure the coherently modulated SP profiles and to verify that the quantum character of the input states is not degraded upon transformation of the mode.

\begin{figure}[ht]
	\includegraphics[width=1.0\linewidth]{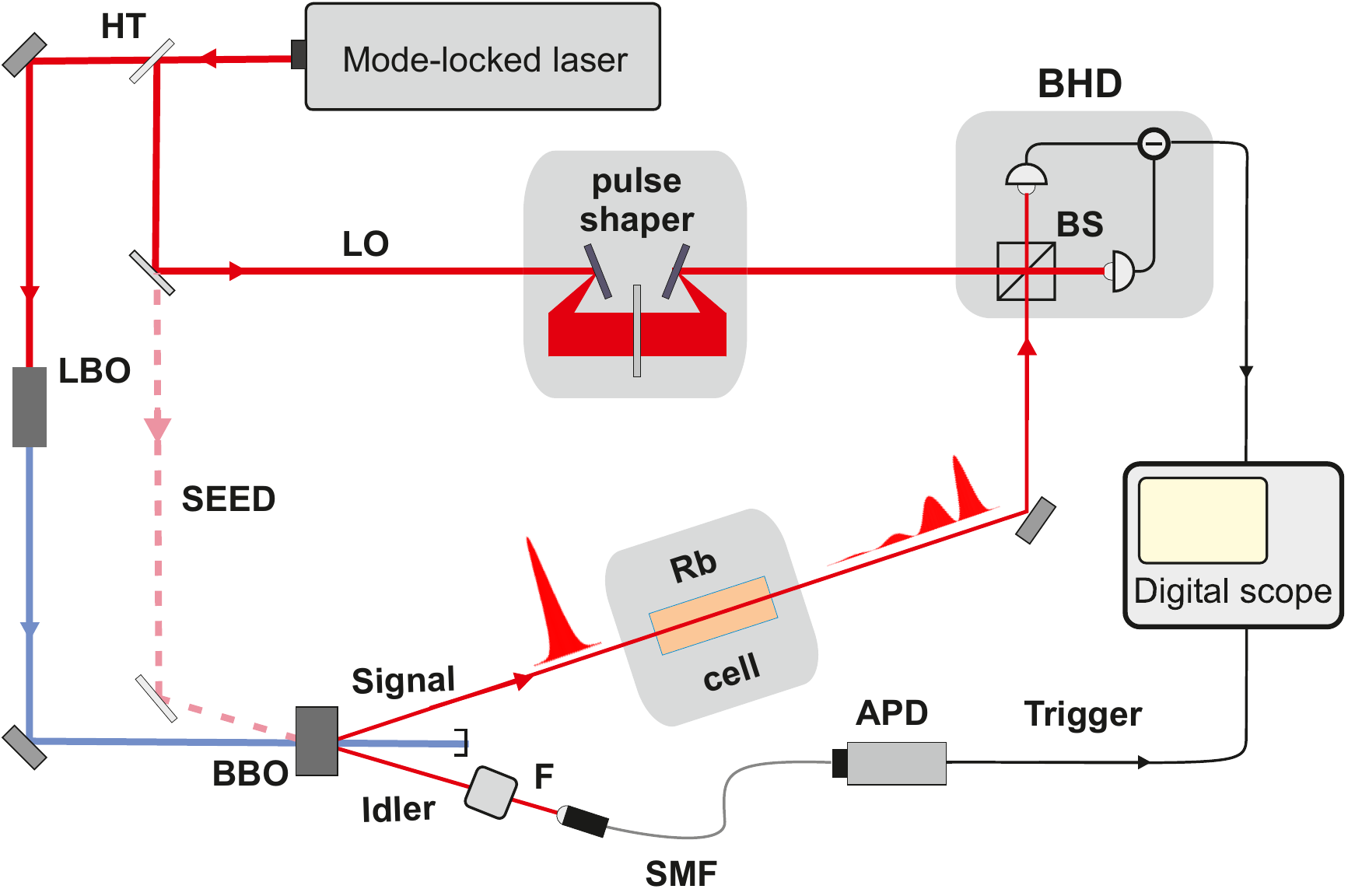}
	\caption{(Color online) Experimental setup. Heralded single-photon pulses interact with a resonant Rb vapor and are analyzed by a balanced homodyne detector (BHD). A seed coherent pulse is used to provide reference classical pulses in the same spatiotemporal mode of the single photons. HT is a high-transmission beam-splitter, LBO is a lithium triborate crystal for second harmonic generation. All other symbols are defined in the text.}
	\label{fig:setup}
\end{figure}
The master light source for our experiment (see Fig.\ref{fig:setup}) is a mode-locked Ti:sapphire laser, emitting a train of 100-fs, 780-nm, pulses at a repetition rate of 80 MHz. Most of the laser emission is frequency-doubled to produce the pump pulses for generating single photons in a traveling wavepacket mode from heralded spontaneous parametric down-conversion (SPDC) in a 300-$\mu$m thick crystal of type-I $\beta$-barium borate (BBO). Frequency-degenerate signal and idler photons are emitted along a cone; the detection of an idler photon in a fiber-coupled avalanche photodiode (APD) placed after narrow spectral and spatial filters (a 1 nm FWHM interference filter, F, and the APD single-mode fiber, SMF) heralds the generation of pure broadband single photons resonant with the $5S_{1/2}-5P_{3/2}$ D2 transition of Rb in the conjugated signal mode. Previous investigations~\cite{Polycarpou2012} have shown that our setup is able to generate SP pulses in a quasi-transform-limited wavepacket mode with a duration of about 100 fs.

A smaller portion (about 8 mW) of the laser emission serves as the local oscillator (LO) for heralded, time-domain, homodyne detection~\cite{Zavatta2002} after mixing with the signal on a 50$\%$ beam-splitter (BS).
When analyzing heralded single photons, we measure a state which can be generally written as
\begin{equation}
\hat \rho_{meas} = \eta \ket{1}_{LO}\!\!\bra{1} +(1-\eta)\ket{0}_{LO}\!\!\bra{0}
\label{eq_eta}
\end{equation}
i.e., a mixture of vacuum and SP Fock states in the mode defined by the homodyne LO. The global efficiency $\eta$ depends upon several factors: limited state preparation efficiency, limited quantum efficiency of the homodyne photodiodes, optical losses, electronic noise, and dark counts in the APD. Above all, however, it depends on the mode mismatch between the heralded state and the LO. Spatial matching is simply obtained by means of lens combinations after pinhole spatial filtering. Spectro-temporal mode-matching can be achieved by shaping the spectral amplitude and phase of the LO by means of a liquid-crystal pixelated spatial light modulator in a 4-f, zero-dispersion, configuration~\cite{Polycarpou2012,Weiner2011}. Using an optimized spatiotemporal LO mode, we routinely measure SP fractions up to $\eta \approx 62\%$ in our setup for the freely propagating conditions.

\begin{figure}[ht]
	\includegraphics[width=1.0\linewidth]{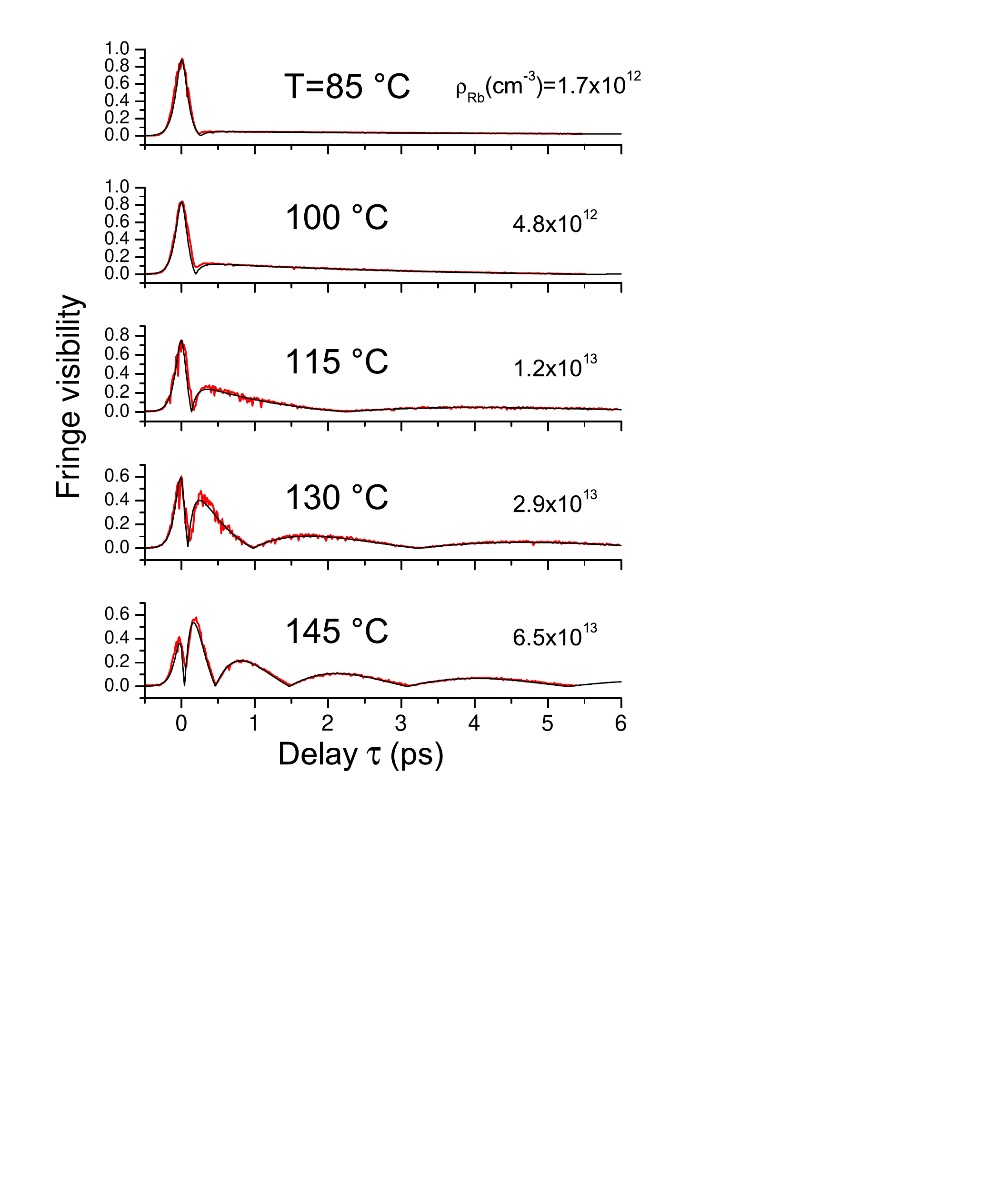}
	\caption{(Color online) Experimental visibility of the interference fringes as a function of the delay (linear cross-correlation) between an un-modulated LO pulse of $\approx100$ fs FWHM duration and classical pulses passing through the resonant Rb cell. At higher temperatures, corresponding to higher estimated Rb densities, the pulses are seen to acquire a typical 0-area shape extending over several picoseconds. Solid black curves are field amplitudes calculated according to Eq.\ref{eq_spmod}, with optical depths $\alpha_0 l= 70, 180, 440, 1000, 2200$ (from top to bottom), and using for $T_2$ the Doppler inhomogeneous lifetimes (between 280 and 260 ps) at the measured temperatures.}
	\label{fig:classical_cc}
\end{figure}

In order to preliminarily test the system and optimize the homodyne detection of modulated SP wavepackets, we first make use of the intense stimulated emission that takes place in the output signal mode of the parametric crystal when seeded by intense synchronized 780 nm pulses in the input idler mode. Emission of this intense classical radiation is known to take place into a mode that closely matches that of the heralded photons~\cite{Aichele2002}. A temperature-stabilized 8-cm long cell containing natural abundance Rb is placed in the common path of the classical and SP wavepackets.
Figure \ref{fig:classical_cc} shows the visibility of the interference fringes measured at one exit of the homodyne beam-splitter between such classical pulses and un-modulated LO pulses as a function of the relative delay for different values of the cell temperature. Starting from a single peak at zero delay, this linear cross-correlation curve develops growing lobes at higher temperatures that clearly testify the formation of classical 0-area pulses. 

The seed idler pulses are then blocked to perform homodyne measurements of the modulated SP wavepackets. When the Rb cell is first inserted in the path of the heralded single photons at room temperature, the only effect is a slight decrease in the global efficiency $\eta$ due to residual losses in the anti-reflection coated cell windows. However, increasing the temperature of the cell has the effect of rapidly degrading the measured global efficiency if the LO mode is left unchanged. In these situations, we expect the SP temporal mode to start being heavily modulated by atomic dispersion, so that the homodyne detector sees larger and larger fractions of vacuum in the original LO mode as the temperature (and the atomic density) increases.

\begin{figure}[ht]
	\includegraphics[width=1.0\linewidth]{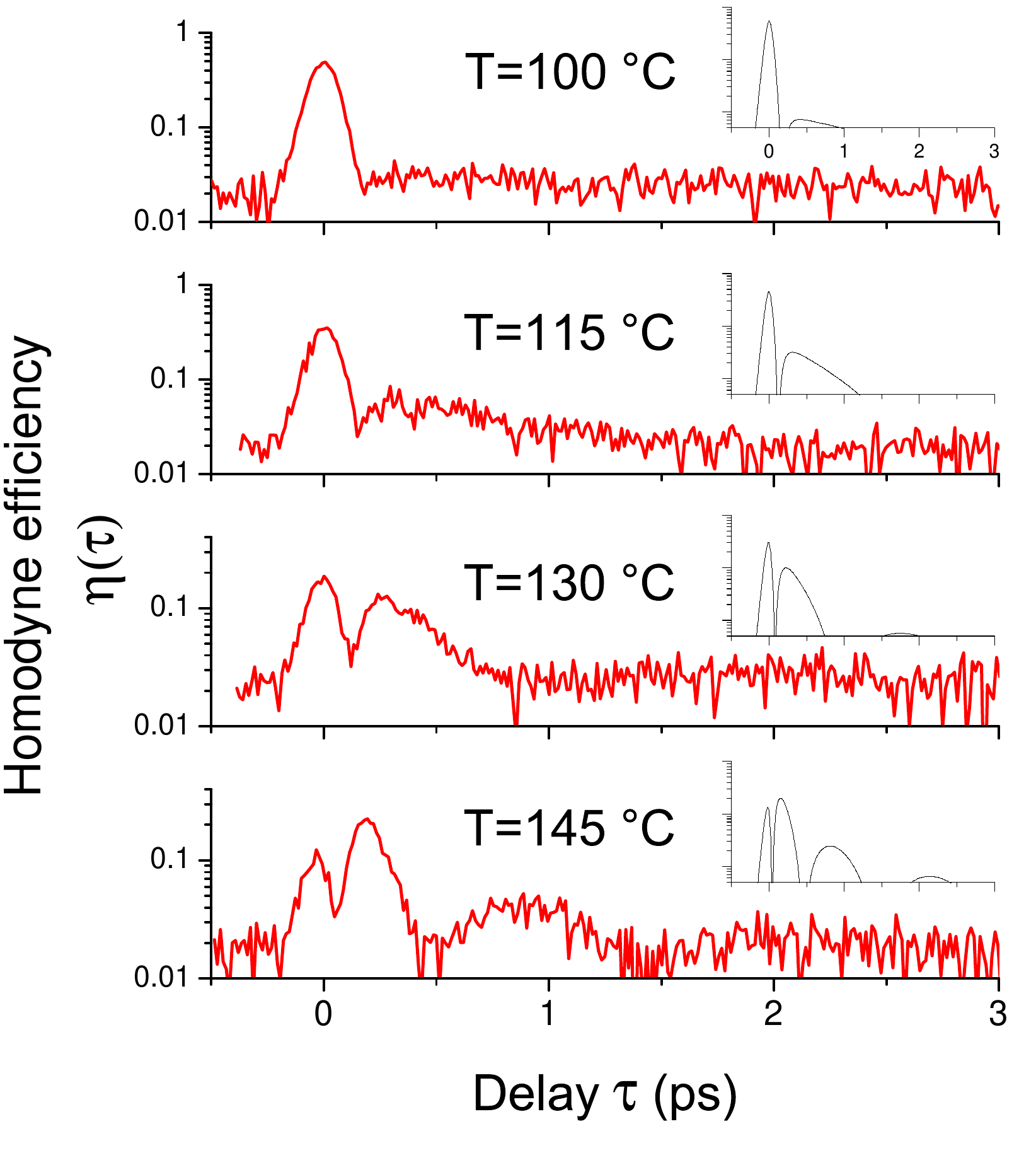}
	\caption{(Color online) Measured homodyne efficiency curves (in logarithmic scale) as a function of the delay between the un-modulated LO pulses and heralded single photons that interacted with the Rb atoms in the cell. The temporal mode of the single photons is heavily distorted at sufficiently high cell temperatures. Theoretical intensity curves calculated with the same parameters of Fig.\ref{fig:classical_cc} are shown in the insets.}
	\label{fig:homo_cc}
\end{figure}

The first series of measurements consists of acquiring phase-averaged quadrature distributions and extracting the global SP efficiency $\eta$ of Eq.\ref{eq_eta} as a function of the delay $\tau$ between the modulated SP pulse and the un-modulated LO. This results in a sort of cross-correlation measurement, similar to the classical one described above, but now performed with the exceptional sensitivity typical of homodyne detection. Figure \ref{fig:homo_cc} shows several plots of $\eta(\tau)$ for different temperatures of the Rb cell. At low temperatures, a single peak of high efficiency is observed, because the LO mode reproduces well the almost unperturbed SP mode. At higher densities, the efficiency curves are seen to acquire strong modulations at increasing delays while the maximum efficiency in the main lobe drops. In these cases, the short LO is no longer well overlapped to the heavily modulated SP, and only a fraction of the latter is detected in the LO mode at any delay. 

\begin{figure}[ht]
	\includegraphics[width=1.0\linewidth]{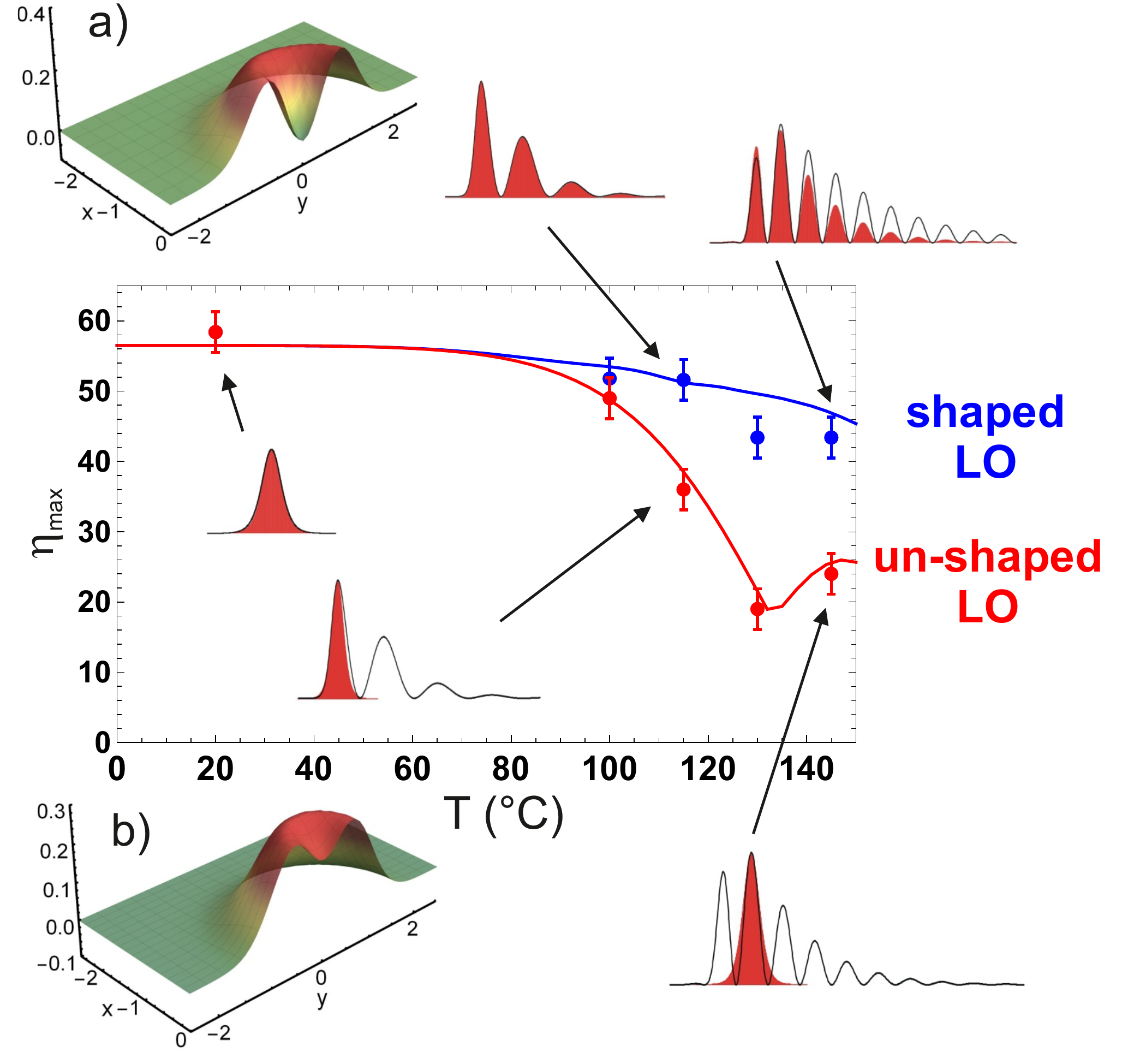}
	\caption{(Color online) Experimental data and theoretical calculations for the maximum homodyne efficiency in the detection of distorted single-photon wavepackets. Red (light gray) points: experimental data for detection with un-modulated LO; blue (dark gray) points: experimental data with optimally shaped LO; red (light gray) and blue (dark gray) solid curves: expected maximum homodyne efficiency in the two cases (including the effect of limited spectral resolution in the LO shaper). a) and b) show the reconstructed Wigner functions of the measured states for the two cases at 115~$^o$C, with and without LO shaping, respectively.}
	\label{fig:eta_vs_t}
\end{figure}

The red (light gray in the grayscale version) experimental data points of Fig.\ref{fig:eta_vs_t} correspond to the maximum efficiencies of these curves, and clearly illustrate the rapid loss of the SP component from the LO mode as the optical depth increases with temperature. The non-monotonous behavior of the data at higher temperatures is easily explained by the fact that, in these conditions, the SP mode gets so distorted that the main pulse actually contains a smaller SP fraction than the first secondary lobe (see last curve of Fig.\ref{fig:homo_cc}).  

Dropping peak efficiencies might suggest that the original, relatively-pure, single photons have been irremediably lost and converted to classical mixtures with vacuum after passing through hot resonant atomic vapors with optical depths in excess of 2000. However, the absorption by the narrow (of the order of GHz) atomic resonance is not to be blamed for this, due to the much larger SP bandwidth (of the order of THz). In fact, the generated single photons are only lost from the homodyne detection mode defined by the un-modulated LO.

In order to recover high homodyne efficiency and verify the generation of highly non-classical 0-area SP wavepackets, one has to measure them in the right mode. This is done by properly shaping the spectrotemporal profile of the reference LO by means of the pulse shaper. A preliminary tuning of the LO shape is performed, for each cell temperature, by maximizing the fringe visibility at zero delay between the shaped LO and the classical 0-area pulses. These shaper settings then serve as the starting point for LO mode optimization in the homodyne setup to detect the shaped single photons. The different measurement scenarios are pictorially illustrated in Fig.\ref{fig:eta_vs_t}, with the different levels of matching between the (shaded) LO shape and the modulated single-photon wavepacket.

The blue (dark gray) data points in Fig.\ref{fig:eta_vs_t} correspond to the peak efficiencies obtained with properly shaped LO modes. It is evident that a matched LO is much more effective in detecting the modulated single photons, and that global efficiencies above 50$\% $, a straightforward proof of their quantum character, with clearly negative Wigner functions (see Fig.\ref{fig:eta_vs_t}a, reconstructed without any correction for limited homodyne efficiency), are recovered also for relatively strong modulations of the SP profile (see the visibility and efficiency curves of Figs.\ref{fig:classical_cc} and \ref{fig:homo_cc}, for 100 and 115~$^o$C).

The residual decreasing trend of the measured SP fraction for increasing temperature is mostly due to the limited spectral resolution of the LO pulse shaping system. Indeed, the current combination of gratings, lenses and liquid-crystal modulators, leads to a minimum spectral resolution of about 0.6 nm, which bounds the range of temporal modulations that it can achieve to a few picoseconds. When the high atomic optical depth causes the single photons to significantly spread over longer delays, the shaped LO mode cannot perfectly overlap any more and the resulting homodyne efficiency decreases. 
By taking the effect of limited spectral resolution of our shaping system into account while calculating the maximum expected homodyne efficiency, we obtain the solid blue (dark gray) curve of Fig.\ref{fig:eta_vs_t}, which reproduces quite well the observed decrease with temperature. The possible distortion of the SP wavefront due to turbulence close to the hot cell surfaces is probably responsible for the additional degradation observed in the experimental data points at the highest temperatures.

\smallskip  
In conclusion, we have demonstrated the re-shaping of the spectrotemporal mode of ultrashort single-photon pulses while propagating through dense resonant atomic vapors. While the single-quantum character of the broadband light states is preserved due to negligible absorption by the narrow atomic transition, the temporal shape of the photon wavepacket is deeply modified, assuming a characteristic 0-area modulated profile, never observed before for non-classical light pulses.
The phenomenon of 0-area classical pulse formation is generally connected to successive interfering cycles of absorptions and stimulated emissions of the pulsed incident radiation by the resonant atomic ensemble~\cite{Dudovich2002}. Its observation at the single-photon level implies an intriguing change of perspective and has interesting links to the phenomenon of single-photon superradiance~\cite{Scully2006} and to new possibilities for the storage of quantum information.

\section{Acknowledgments}
This work was partially supported by the EU under the ERA-NET CHIST-ERA project QSCALE, and by the Brazilian agencies CNPq and CAPES.

\end{document}